\input epsf 
\documentstyle[preprint,aps,eqsecnum,floats,epsf]{revtex}
\begin{document}

\draft
\preprint{\vbox{
                 \hfill      DOE/ER/40762--072 \\
                 \null\hfill UMPP \#96--056 \\
                 \null\hfill AZPH-TH-95/29
               }
         }

\date{Original, December 5, 1995; revised March 29, 1996} 

\title
{Photon Contribution to the ``Super-hard" Pomeron}
\author{Hung Jung Lu}
\address
{Department of Physics,  University of Arizona \\
Tucson, Arizona 85721, U.S.A.}
\author{Joseph Milana}
\address
{Department of Physics, University of Maryland \\
College Park, Maryland 20742, U.S.A.}
\maketitle

\vspace{0.3in}

\begin{abstract}\widetext
It is shown that  direct photons provide  a leading twist mechanism  for 
diffractive jet production in which the jets carry away 
all of the momentum lost by the proton.   Two-photon 
processes are thus expected to asymptotically dominate 
 ``super-hard'' pomeron events in $ep$ collisions. We  
report the expected rates from these events for  recent  
ZEUS and H1 data cuts.    We also estimate the direct photon 
contribution to the ``super-hard" pomeron events 
observed by the CERN UA8 group for $p\bar p$  collisions.  
It is again shown that direct photons are the leading twist mechanism 
for these events.    We find that direct photons are an appreciable 
fraction of the events seen by UA8.
\end{abstract}
\vglue0.25in
\narrowtext

\newpage
\section{Introduction}

Jet events with large rapidity gaps in the forward proton
direction have been observed at the DESY HERA $ep$ collider,
both in the deep-inelastic regime \cite{ZeusDIS,H1DIS} and in 
the photoproduction regime \cite{ZeusPhotoproduction,H1Photoproduction}.
In the now popular convolution picture of the pomeron\cite{IngelmanSchlein}, 
the virtual or quasi-real photon emitted from the electron collides
with the partons inside the pomeron, leading to the
production of hard jets, well-separated in the phase
space from the forward-going proton.   For the diffractive
photoproduction events, Collins, Frankfurt and Strikman
\cite{CollinsFrankfurtStrikman} have argued that a special subclass of
diffractive events 
would be produced at higher twist in the jet-system
invariant mass due to the color singlet structure of the dijet system.   
These are events where the pomeron transfers all of its energy
into the hard-jet system.    Such events are strictly impossible in 
convolution pictures due to the assumed, normal vanishing of parton
structure functions 
 at $x \rightarrow 1$.    These predictions, restated in Ref.
\cite{BerSop}, are explicitly 
manifested in models\cite{LuMilana,DL2} where these events have been generated 
assuming  a two gluon substructure for the exchanged 
pomeron.\footnote{The possibility\cite{LuMilana} of differing 
small--$x$ behavior for the 
superhard and ``normal'' pomeron type events may however complicate testing 
these predictions.}

Events where the initial proton scatters off diffractively
but generates hard-jet systems in the central rapidity
region are also known as diffractive hard-jet events, or
simply, hard diffraction.
These events have been previously observed at $p\bar p$
collision. The UA8 experiment at the
CERN $Sp\bar{p}S$ Collider \cite{UA8} has studied
the production of jet event in the single diffractive 
regime. That is, 
\begin{equation}
p + \bar{p} \to p  + ({\rm jets} + X),
\label{EqUA8Reaction}
\end{equation}
or its conjugate reaction (reverse the roles
of $p$ and $\bar{p}$). 
In the above process, the incident antiproton 
interacts with a soft pomeron component of the proton 
to generate the hard-jets, while the diffracted proton 
preserves most ( $> 90 \%$ ) of its initial beam energy. 
It is observed in this experiment that an
unexpected large fraction of the pomeron's
momentum participates in the hard scattering
a significant fraction of the time.   The
hypothetical pomeron therefore seems to
contain a point-like component, capable of
transferring all its energy into the
hard jet system.    The predictions of Collins, Frankfurt and 
Strikman\cite{CollinsFrankfurtStrikman} is 
hence that factorization is violated at the leading twist level in hard
diffractive processes.  
More data, determining the superhard pomeron cross--section as a function of  
  the transverse momentum $P_\perp$ of the jets, is however  still 
needed before these interpretations of the UA8 results are properly confirmed.
\footnote{What is at issue is the role of Sudakov effects\cite{Mueller} 
known to play a crucial role in elastic scattering events at large momentum 
transfers\cite{BottsSter}.    See section III(a) below.  
Indeed it is due to these Sudakov effects that it was argued 
in Ref. \cite{LuM} that exclusive hard double diffractive events would
be higher twist,  in contradistinction to the proposed mechanisms of 
Refs. \cite{BialasLandshoff,Pumplin,BerCollins}.}

Here we study the contribution of direct 
photon from the proton to the above hard-jet
events at DESY HERA and CERN $Sp\bar{p}S$ colliders.
Although this contribution is suppressed 
by the electromagnetic coupling,
it provides a mechanism that is leading twist 
in the jet-pair invariant mass.  Hence in the case of $ep$ collisions, it is 
expected to asymptotically  dominate all hadronic mediated dijet events
in which the 
jets carry away all of the momentum lost by the proton. 
We find that at the present energy and luminosity level,
the two-photon contribution accounts for
about 5\% of the observed large rapidity gap, two-jet event cross 
section recently reported by the ZEUS\cite{ZeusPhotoproduction} and 
H1\cite{H1Photoproduction} collaborations at HERA 
(assuming 100\% detector efficiencies).    It is however worth stressing that 
in both experiments the final state proton's momentum was not determined   
 and hence  what fraction of these  diffractive events involved a total transfer of 
the proton's momentum to the two jets could not be determined.

In the case of the $p\bar{p}$ events of Eq.~(\ref{EqUA8Reaction}),
direct photons are again the leading twist mechanism.   
Due to the failure of factorization of 
the hadronic mechanism, this fact requires a separate argument to demonstrate.  
Nevertheless the conclusion that asymptotically, direct photons dominate 
is once again valid.    In application to realistic energies, we find  
 that direct photon's contribute $~15$\% of the events observed at UA8.   
While this result indicates that the (higher--twist) hadronic mechanisms
are still dominant, direct photons are clearly relevant.  The two mechanisms 
(strong vs. electromagnetic) may in fact be quite comparable  as their 
contributions must be added to obtain the $S$--matrix. 

\section{Two-Photon Process Jet Events at HERA}
\begin{figure}[htbp]
\begin{center}
\leavevmode
{
 \epsfxsize=6.00in
 \epsfbox{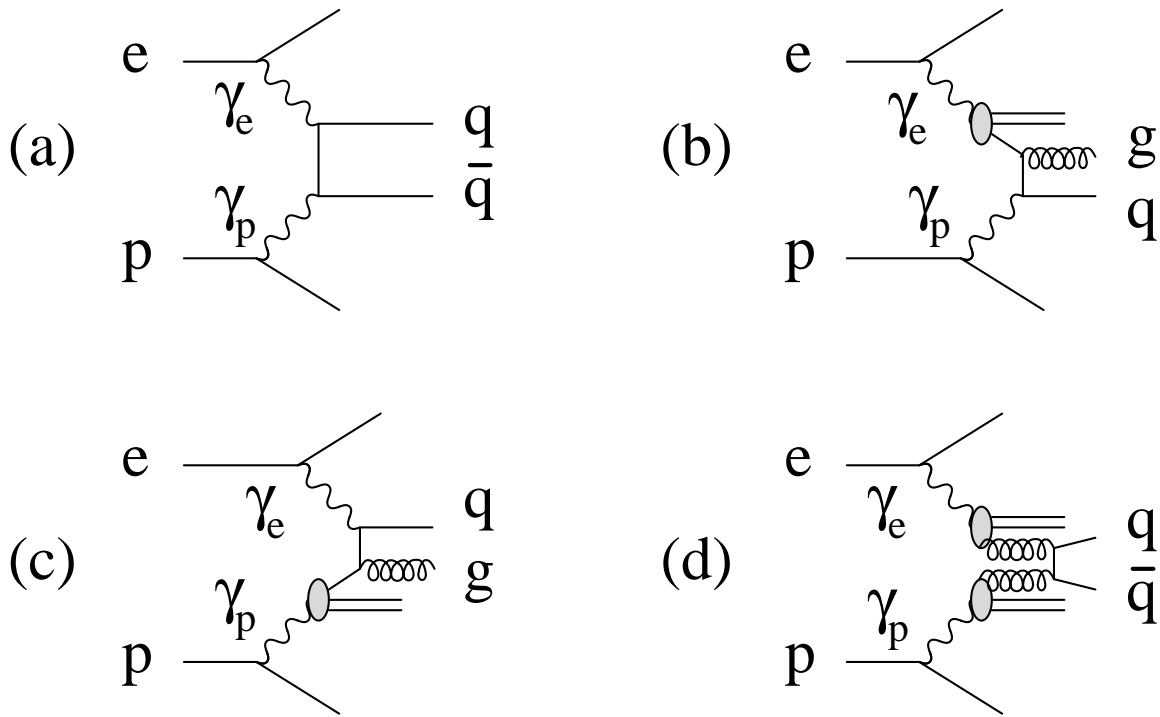}
}
\end{center}
\caption[*]{
           Two-photon jet processes at HERA.
           (a) direct contribution, 
           (b) resolved $\gamma_e$ contribution, 
           (c) resolved $\gamma_p$ contribution, 
           (d) twice-resolved contribution. 
           }
\label{Fig1}
\end{figure}

In this section we analyze the two-photon mechanisms
and their contribution to the generation of jet events
at HERA in the diffractive regime. The scattering 
mechanisms are illustrated in Fig. 1, where we have
only shown some representative diagrams. The actual
number of diagrams is large, especially for the
twice resolved mechanism of Fig. 1 (d). 

For the two-photon processes, quasi-real 
photons are generated from the incoming electron and proton. 
The parent beam particles scatter off diffractively, 
leaving the photons to interact and generate 
the hard jet event in the central-rapidity region.
The underlying photons ($\gamma_e, \gamma_p$) can interact 
either directly, as in Fig. 1 (a), or via the resolved photon 
mechanisms (see \cite{DreesGodbolePramana} and references
there-in), as in Fig. 1 (b), (c) and (d). The processes
(a) and (b) are to be compared to the emission of a 
``super-hard" pomeron from the proton, where all the energy
lost by the proton has gone into the hard jet system.

The photon content of the electron and of the proton
can be adequately described by the equivalent photon 
formalism (EPA)\cite{WeizsackerWilliams}. This $ep$ analogy 
of the two-photon mechanism at $e^+e^-$ colliders 
\cite{DreesGodbolePramana,EETwoPhoton}
has been pointed out in Ref. \cite{Bussey}.
For the photon distribution inside the electron, 
we use the expression
\begin{equation}
f_{\gamma|e}(x_1)=
\frac{\alpha_{\rm em}}{2\pi}
\left[
      \ln \left(  
                \frac{1-x_1}
                     {x_1^2}
                \frac{ Q^2_{~\rm max} }
                     {m_e^2}
          \right)
      \frac{ 1 + (1-x_1)^2 }{x_1} 
    - \frac{2(1-x_1)}{x_1} 
\right],
\end{equation}
where $x_1$ is the momentum fraction of the electron
carried by its emitted photon, 
$\alpha_{\rm em} \sim 1/137$ the QED coupling constant,
$m_e$ the mass of the electron, and
$Q^2_{\rm max}$ is the maximum 
transverse momentum squared as allowed by the
experimental cut. For instance, in the case of HERA ZEUS 
experiment \cite{ZeusPhotoproduction}, the diffractive events
are selected with 
$Q^2_{\rm max} \sim 4$ GeV$^2$.

For the proton, the derivation of the equivalent photon
formula mimics the electron case, except that
now one has to take into account the Dirac ($F_1$) and Pauli 
($F_2$) form factors,  
which appear in the photon--proton vertex as
\begin{equation}
\bar{u} \Lambda_\mu u = i e \bar{u} \left[ \gamma_\mu F_1(q^2) + \frac{i}{2 m_p} 
\sigma_{\mu \nu} q ^\nu F_2(q^2) \right] u,
\end{equation}
with $\sigma_{\mu \nu} = \frac{i}{2}\left[ \gamma_\mu, \gamma_\nu
\right]$ and $q^\nu$ the 
four--momentum flowing into the vertex.  
(Note: EPA for heavy nuclei have studied in \cite{GamGamHiggs}.)  
The resulting photon distribution is
\begin{eqnarray}
f_{\gamma|p}(x_2)
&=&
\frac{\alpha_{\rm em}}
     {2\pi}
\int_{x_2^2 m_p^2 /(1-x_2)}^\infty
\frac{d Q^2}
     { Q^2 }
\left[ 2 \left( F_1^2(Q^2) + \frac{Q^2}{4 m_p^2} F_2^2(Q^2) 
\right) \times \right.\\
\nonumber
&&\left. \left(  \frac{1-x_2 }{x_2}-  \frac{m_p^2}{Q^2} x_2 \right)
+ \left( F_1(Q^2) + F_2(Q^2) \right)^2 x_2  \right].
\label{fphinprot}
\end{eqnarray}
Using as experimental input that the electric and
magnetic form factors are well described
as dipoles\cite{ffdata}
($G_E(Q^2) \approx G_D(Q^2) \approx G_M(Q^2)/\mu_p$),
one obtains that
\begin{eqnarray}
F_1(Q^2) &=& G_D(Q^2) \frac{ 1 + \frac{\mu_p Q^2}{4 m^2} }
{ 1 + \frac{Q^2}{4 m^2}}\\
\nonumber
F_2(Q^2)&=& G_D(Q^2) \frac{\mu_p - 1}{ 1 + \frac{Q^2}{4 m^2}}\\
\nonumber
G_D(Q^2) &=& \frac{1}{(1+Q^2/Q_0^2)^2}
\end{eqnarray}
with $m_p \sim .938$ GeV, $Q_0^2 \sim 0.71$ GeV$^2$, $\mu_p=2.7928$  and
$Q^2 = |t| \sim p_{\rm t}^2$ the momentum transfer
squared of the proton carried by its photon.
In the small-$x$ region, the photon's distribution is roughly given by
\begin{equation}
f_{\gamma|p}(x_2)
\sim
\frac{\alpha_{\rm em}}
     {\pi x_2}
\left[
      2 \ln \left(
                \frac{1}{x_2}
          \right)
      - 1.627
\right].
\end{equation}
The large logarithm arises from the Dirac form factor and in practice
the Pauli form factor 
is ignorable for either very low (e.g. the HERA data to be discussed
below) or very high $Q^2$.\footnote{Similarly, resonance excitation such as 
the $\Delta(1232)$ which couples to the proton through a Pauli type
vertex\cite{Nimai} is 
likewise ignorable.}  The UA8 data to be discussed in section III falls
into neither of these 
categories.

The differential cross section for the two-jet production
mechanism as shown in Fig. 1 is given by
\begin{equation}
d \sigma_{ep}
=
f_{\gamma|e} (x_1)
f_{\gamma|p} (x_2)
d x_1 d x_2
d \sigma_{\gamma\gamma}.
\label{cross}
\end{equation}

The $\gamma\gamma \to 2 \ jets$ cross section receives
contribution from all the mechanisms shown in Fig. 1,
\begin{equation}
  d \sigma_{\gamma\gamma}
= d \sigma_{\gamma\gamma}^{(a)}
+ d \sigma_{\gamma\gamma}^{(b)}
+ d \sigma_{\gamma\gamma}^{(c)}
+ d \sigma_{\gamma\gamma}^{(d)}.
\end{equation}
where $d \sigma_{\gamma\gamma}^{(a)}$ is the direct photon
contribution, $d \sigma_{\gamma\gamma}^{(b)}$ is the resolved
$\gamma_e$ contribution, $d \sigma_{\gamma\gamma}^{(c)}$ is 
the resolved $\gamma_p$ contribution, and 
$d \sigma_{\gamma\gamma}^{(d)}$ is the ``twice-resolved"
contribution.

\subsection{Direct Photons are Leading Twist}

The direct photon contribution consists of 
the $\gamma\gamma \to q \bar{q}$ differential cross section
\begin{equation}
d \sigma_{\gamma\gamma}^{(a)} 
=
\frac{12 \pi \alpha_{\rm em}^2}
     {x_1 x_2 s}
\left(
     \sum_q Q_q^4
\right)
\frac{1+z^2}
     {1-z^2}
\ d z,
\label{direct}
\end{equation}
where we have summed over the final quark colors.
$Q_q$ is the electric charge of the quark of flavor $q$,
$\sum Q_q^4 = 34/81$ for 4 light-quark flavors ($u,d,s,c$). 
We have used the integration variable $z=\cos\theta$ 
with the angle $\theta$ being the scattering angle in the 
parton center-of-mass system. The total center-of-mass 
energy squared of the $ep$ system is $s=4 E_e E_p$. 
Currently at HERA $E_e=26.7$ GeV and
$E_p=820$ GeV, thus $\sqrt{s}=296$ GeV.  

Observe that the contribution of the direct photon is leading twist.  
Inserting Eq.~(\ref{direct}) into Eq.~(\ref{cross}) one finds that the contribution 
of direct photons to the cross--section is parametrically 
\begin{equation}
d \sigma_{ep}^{(a)}
\propto
\frac{\alpha_{\rm em}^3}
     {x_1 x_2 s}
\frac{1+z^2}
     {1-z^2}
\ d z \frac{ d x_2}{x_2} f_{\gamma|e} (x_1) d x_1 
\end{equation}
where we have focussed on the leading $x_2^{-1}$ behavior of 
$f_{\gamma|p} (x_2)$.  
The contribution of a hadronic mediated pomeron is  
higher--twist for this class of dijet events.  
 Parametrically, two  groups modeling the pomeron with a two gluon substructure    
find\cite{LuMilana,DL2} that the hadronic, superhard pomeron dijet 
cross--section behaves as
\begin{equation} 
d \sigma_{ep}^{gg}
\propto
\frac{1}{b P_\perp^2} \frac{\alpha_{\rm em} \alpha_{\rm s}^2 }
     {x_1 x_2 s}
\frac{1+z^2}
     {1-z^2}
\ d z \frac{ d x_2}{x_2} f_{\gamma|e} (x_1) d x_1, 
\end{equation}
where $P_\perp$ is the transverse momentum of each jet and $b \approx 4$
GeV$^{-2}$ enters in the proton--pomeron's hadronic form factor, $e^{bt}$.     

The necessity of adding an extra gluon to the color singlet, dijet system has 
resulted in the hadronic mechanism being higher twist.  
Asymptotically ($P_\perp^2 \rightarrow \infty$) 
the direct-photon mechanism dominates in this class of 
diffractive events.  

\subsection{Resolved Photon Contributions}
The resolved $\gamma_e$ cross section is given by
(taking into account the quark-antiquark symmetry
in the parton distribution function inside photon)
\begin{equation}
d \sigma_{\gamma\gamma}^{(b)}
=
\left[
2 \sum_q f_{q|\gamma}(y_1,p^2_{\rm t}) d \sigma_{q\gamma}(x_1 x_2 y_1 s)
+ f_{g|\gamma}(y_1,p^2_{\rm t}) d \sigma_{g\gamma}(x_1 x_2 y_1 s)
\right] 
d y_1,
\end{equation}
where $y_1$ is the momentum fraction of $\gamma_e$ carried
by its parton, $f_{q|\gamma}(y_1,p^2_{\rm t})$ and 
$f_{g|\gamma}(y_1,p^2_{\rm t})$ are the parton distribution
functions inside photon, where we have used the transverse
momentum squared $p^2_{\rm t}=x_1 x_2 y_1 (1-z^2) s /4 $
as the factorization scale. The parton-photon cross sections are 
\begin{eqnarray}
d \sigma_{q\gamma} (\hat{s}) 
&=&
\frac{2\pi}{3 \hat{s}}
Q_q^2
\alpha_{\rm em}
\alpha_{\rm s}
      \frac{5+2 z+z^2}{1+z}
\ d z,
\label{QuarkPhotonCrossSection}
\\
d \sigma_{g\gamma} (\hat{s})
&=&
\frac{2\pi}{\hat{s}}
\left( 
       \sum_q Q_q^2
\right)
\alpha_{\rm em}
\alpha_{\rm s}
      \frac{1+z^2}
           {1-z^2}
\ d z.
\label{GluonPhotonCrossSection}
\end{eqnarray}
Similarly, the resolved $\gamma_p$ contribution is given by 
\begin{equation}
d \sigma_{\gamma\gamma}^{(c)}
=
\left[
2 \sum_q f_{q|\gamma}(y_2,p^2_{\rm t}) d \sigma_{q\gamma}(x_1 x_2 y_2 s)
+ f_{g|\gamma}(y_2,p^2_{\rm t}) d \sigma_{g\gamma}(x_1 x_2 y_2 s)
\right] 
d y_2,
\end{equation}
with $y_2$ the momentum fraction of $\gamma_p$ carried by
its parton, and $p^2_{\rm t} =x_1 x_2 y_2 (1-z^2) s /4 $
the transverse momentum squared.

The twice-resolved cross section $d \sigma_{\gamma\gamma}^{(c)}$
is substantially more complicated that the previous cases.
The underlying parton $2 \to 2$ matrix elements  
can be found in Ref. \cite{BargerPhillips}. Taking
into account the quark-antiquark symmetry of the parton
distribution functions inside photon, we have
\begin{eqnarray}
d \sigma_{\gamma\gamma}^{(d)}
=
\frac{\pi\alpha_{\rm s}^2}
     {2 x_1 x_2 y_1 y_2 s}
&&
\left\{
       4 \sum_{q \ne q'} 
              f_{q|\gamma}(y_1,p_{\rm t}^2)
              f_{q'|\gamma}(y_2,p_{\rm t}^2)
       {\cal M}^2_{\rm I}
\right.
\nonumber
\\
&&
+ 2 \sum_{q}
    f_{q|\gamma}(y_1,p_{\rm t}^2)
    f_{q|\gamma}(y_2,p_{\rm t}^2)
\left[
      {\cal M}^2_{\rm II}
    + (N_f-1) {\cal M}^2_{\rm III}
    + {\cal M}^2_{\rm IV}
    + {\cal M}^2_{\rm V}
\right]
\nonumber
\\
&&
+
f_{g|\gamma}(y_1,p_{\rm t}^2)
f_{g|\gamma}(y_2,p_{\rm t}^2)
\left[
      4 {\cal M}^2_{\rm VI}
      + {\cal M}^2_{\rm VIII}
\right]
\\
&&
\left.
+ 2 \sum_{q}
    \left[
          f_{q|\gamma}(y_1,p_{\rm t}^2)
          f_{g|\gamma}(y_2,p_{\rm t}^2)
        + f_{g|\gamma}(y_1,p_{\rm t}^2)
          f_{q|\gamma}(y_2,p_{\rm t}^2)
    \right]
{\cal M}^2_{\rm VII}
\right\}
\ d y_1 d y_2 d z.
\nonumber
\end{eqnarray}
In the previous expression, $N_f$ is the number of light-quark
flavors, $p_{\rm t}^2 = x_1 x_2 y_1 y_2 (1-z^2) s / 4$ is the transverse
momentum squared, and the amplitudes are given by\cite{BargerPhillips} 
\begin{eqnarray}
{\cal M}^2_{\rm I}
&=&
\frac{4}{9} \
\frac{\hat{s}^2+\hat{u}^2}{\hat{t}^2}
\nonumber\\
{\cal M}^2_{\rm II}
&=&
\frac{2}{9}
\left(
      \frac{\hat{s}^2+\hat{u}^2}{\hat{t}^2}
    + \frac{\hat{s}^2+\hat{t}^2}{\hat{u}^2}
\right)
-\frac{4}{27}
\frac{\hat{s}^2}{\hat{u}\hat{t}}
\nonumber\\
{\cal M}^2_{\rm III}
&=&
\frac{4}{9} \
\frac{\hat{t}^2+\hat{u}^2}{\hat{s}^2}
\nonumber\\
{\cal M}^2_{\rm IV}
&=&
\frac{4}{9}
\left(
      \frac{\hat{s}^2+\hat{u}^2}{\hat{t}^2}
    + \frac{\hat{t}^2+\hat{u}^2}{\hat{s}^2}
\right)
-\frac{8}{27}
\frac{\hat{u}^2}{\hat{s}\hat{t}}
\nonumber\\
{\cal M}^2_{\rm V}
&=&
\frac{16}{27}
\frac{\hat{u}^2+\hat{t}^2}{\hat{u}\hat{t}}
-\frac{4}{3}
\frac{\hat{u}^2+\hat{t}^2}{\hat{s}^2}
\nonumber\\
{\cal M}^2_{\rm VI}
&=&
\frac{1}{6}
\frac{\hat{u}^2+\hat{t}^2}{\hat{u}\hat{t}}
-\frac{3}{8}
\frac{\hat{u}^2+\hat{t}^2}{\hat{s}^2}
\nonumber\\
{\cal M}^2_{\rm VII}
&=&
\frac{\hat{s}^2+\hat{u}^2}{\hat{t}^2}
-\frac{4}{9}
\frac{\hat{s}^2+\hat{u}^2}{\hat{u}\hat{s}}
\nonumber\\
{\cal M}^2_{\rm VIII}
&=&
\frac{9}{8}
\left(
      \frac{\hat{s}^2+\hat{u}^2}{\hat{t}^2}
    + \frac{\hat{s}^2+\hat{t}^2}{\hat{u}^2}
    + \frac{\hat{u}^2+\hat{t}^2}{\hat{s}^2}
    + 3
\right),
\end{eqnarray} 
where $\hat{t}=-(1-z)\hat{s}/2$, $\hat{u}=-(1+z)\hat{s}/2$,
and $z=\cos\theta$, with $\theta$ the scattering angle in
the parton center-of-mass system. We have multiplied a
factor $1/2$ to the amplitudes involving identical
particle final states.

\subsection{Results}
At HERA, the existence of a large rapidity gap in the forward
(proton) direction has been employed\cite{ZeusPhotoproduction,H1Photoproduction} 
as a selection criterion for diffractive events.   
The diffracted proton's momentum is however not determined 
and thus no information resembling a ``structure'' function decomposition of 
the pomeron {\it a la} UA8\cite{UA8} is available.    The relative number of 
superhard pomeron events, where all the momentum lost by the proton is 
carried away by the jets is thus unknown.  

The large rapidity gap criterion is used because in general one expects
a gap of size $\sim \ln(1/x_2)$ separating the hadron fragments
from the proton, although the actual size of the rapidity gap 
varies from event to event, and depends on the details
of the hadronization physics. For $d \sigma_{\gamma\gamma}^{(a)}$
and  $d \sigma_{\gamma\gamma}^{(b)}$, we expect most hadron 
fragments to be produced in phase space region well separated
from the forward proton. However, for $d \sigma_{\gamma\gamma}^{(c)}$
and $d \sigma_{\gamma\gamma}^{(d)}$, since the forward going $\gamma_p$
is broken, only a fraction of these events will contain a large
rapidity gap between the diffracted proton and the jet hadron fragments.
The precise fraction depends on the details of the hadronization
process. Here we will limit ourselves to the perturbative results, and
keep in mind that the gap event cross section is somewhere between
$d \sigma_{\gamma\gamma}^{(a)}+d \sigma_{\gamma\gamma}^{(b)}$
and
$d \sigma_{\gamma\gamma}^{(a)}
+d \sigma_{\gamma\gamma}^{(b)}
+d \sigma_{\gamma\gamma}^{(c)}
+d \sigma_{\gamma\gamma}^{(d)}$.

Experimentally, in order to observe well-defined jet,
a minimum transverse momentum cut ${p_{\rm t}}_{\rm min}$
is introduced. Also, an additional cut in rapidity is
used at HERA in order to separate the diffractive
events from the proton dissociation events. We will
therefore also consider a maximum rapidity cut 
$\eta_{\rm max}$ in the forward direction (proton's
direction) for the jet rapidities. The laboratory-frame
rapidities of $q$ and $\bar{q}$ jets are
\begin{equation}
\eta_{\rm \ jet}
=
\frac{1}{2}
\ln\left(
         \frac{E_p}{E_e}
   \right)
+
\frac{1}{2}
\ln\left(
         \frac{x_2 y_2}{x_1 y_1}
   \right)
\pm
   \ln\tan\left(
              \frac{\theta}{2}
          \right),
\end{equation}
where for the cases (b), (c) and (d) we take
$y_2=1$, $y_1=1$ and $y_1=y_2=1$, respectively.
Experimentally, detector limit also imposes cuts 
in the momentum fraction $x_1$ carried by $\gamma_e$.
Summarizing, the cross section is subjected to the constraints
\begin{equation}
\left\{
       \begin{array}{c}
              \eta_{\rm \ jet} 
            =
             \frac{1}{2}
             \ln\left(
                      \frac{E_p}{E_e}
                \right)
             +
             \frac{1}{2}
             \ln\left(
                     \frac{x_2 y_2}{x_1 y_1}
                \right)
             \pm 
             \frac{1}{2}
                  \ln\left(
                          \frac{1+z}{1-z}
                     \right) 
             < 
             \eta_{\rm max},
       \\
             p_{\rm t}^2
             =
             \frac{1}{4}
             x_1 x_2 y_1 y_2 s (1-z^2) 
             >
             {p_{\rm t}^2}_{\rm min},
       \\
             {x_1}_{\rm min}  < x_1 < {x_1}_{\rm max},
       \\
             Q^2_{\rm min} < Q^2 < Q^2_{\rm max}.   
       \end{array}
\right.
\end{equation}
Table 1 gives the kinematic constraints as used by
the HERA ZEUS \cite{ZeusPhotoproduction} and H1
\cite{H1Photoproduction} groups. In Fig. 2 we
plot the obtained cross section as function of 
$p_{\rm t}$. The solid lines represent the
contributions from (a) and (b) (unbroken $\gamma_p$), 
and the dotted lines represent the combined contributions
from (a), (b), (c) and (d). The integrated cross section
between $p_{\rm t} = 4 $ GeV and $p_{\rm t} = 12 $ GeV 
are given in Table 2.
For ZEUS we have $\sigma(a+b) = 10.3$ [pb] and 
$\sigma(a+b+c+d) = 13.7$ [pb], and for H1 
$\sigma(a+b) = 3.2$ [pb] and $\sigma(a+b+c+d) = 4.5$ [pb].
At ZEUS, the integrated luminosity is $0.55$ [pb$^{-1}$],
and $132$ two-jet events have been observed with the
given kinematic cuts. At H1, the integrated luminosity
is $0.289$  [pb$^{-1}$], and $19$ two-jet events have been 
observed with the corresponding kinematic cuts. Although
no information on detector efficiency is given in
\cite{ZeusPhotoproduction,H1Photoproduction}, it appears
that the two-photon mechanism contributes to the observed
large-gap two-jet events at about the $5 \%$ level.   
However, we remind the reader that experimentally no detailed 
momentum decomposition of these dijet events is available and the  
``normal'' Ingelman-Schlein\cite{IngelmanSchlein} type of process 
with broken proton final state may have substantial contribution. 
The most relevant rate for our discussion, the superhard 
pomeron type events where the diffractive proton is unaccompanied by any 
additional beam jet hadrons, has not been measured.  
 
\begin{table}[t]
\caption{Experimental kinematic cuts used at HERA ZEUS and
         H1 groups}
\begin{tabular}{|c|c|c|} 
\null             & ZEUS    &  H1   \\ \hline
$\eta_{\rm max}$  & $1.5$   & $1.5$ \\ \hline
${p_{\rm t}}_{\rm min}$  & $4$ GeV & $4$ GeV \\ \hline
$Q^2_{\rm min}$ & $0$ GeV$^2$  & $3 \times 10^{-8}$ GeV$^2$ \\ \hline
$Q^2_{\rm max}$ & $4$ GeV$^2$  & $10^{-2}$ GeV$^2$ \\ \hline
${x_1}_{\rm min}$ & $0.05$ & $0.25$ \\ \hline
${x_1}_{\rm max}$ & $0.80$ & $0.70$ 
\end{tabular}
\end{table}

\begin{figure}[htbp]
\begin{center}
\leavevmode
{
 \epsfysize=8.00in
 \epsfbox{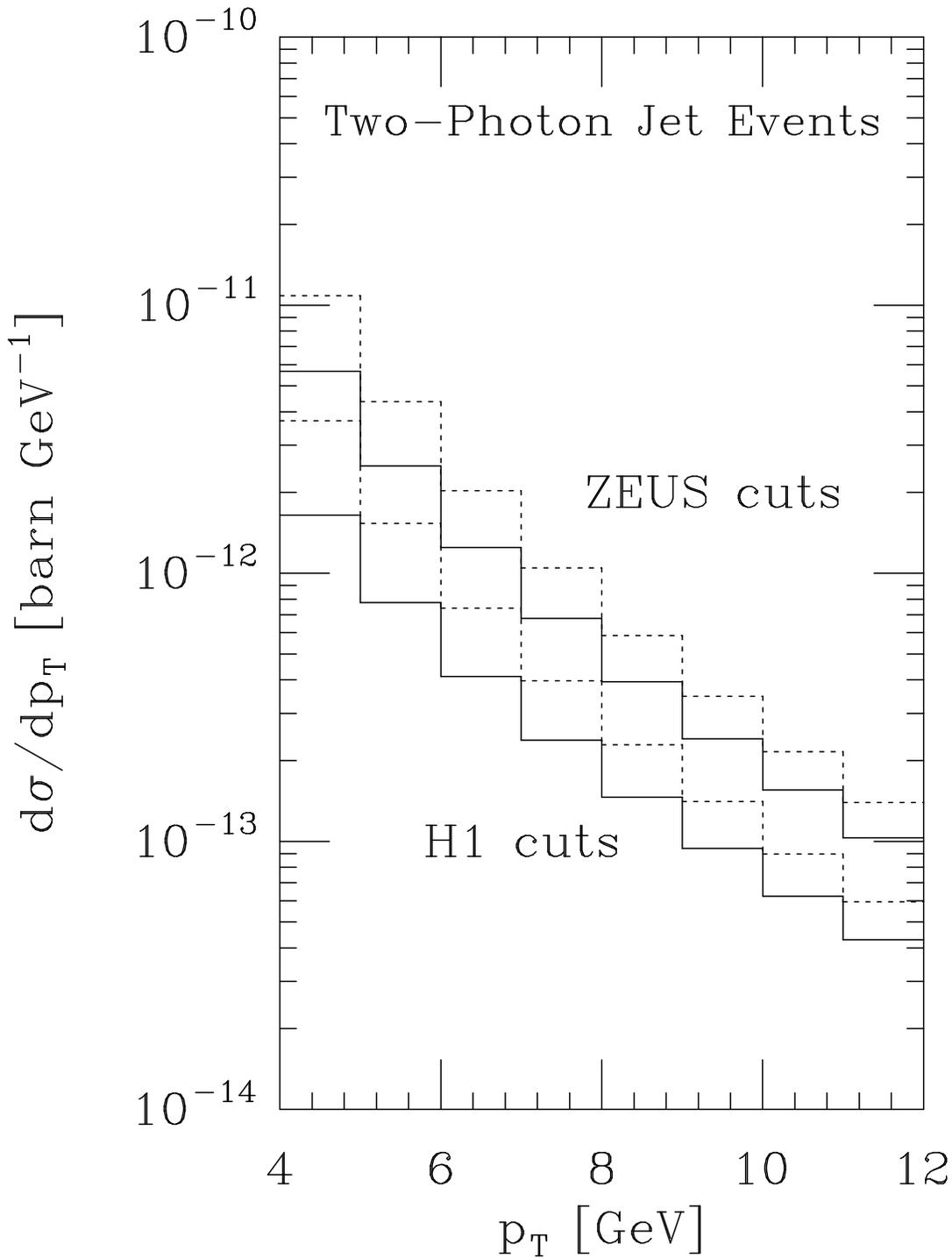}
}
\end{center}
\caption[*]{
           Two-photon jet event cross section, according
           to the ZEUS and H1 kinematic cuts given in
           Table I. The solid lines are obtained by
           considering only the direct and resolved $\gamma_e$
           contribution (a+b). The dotted lines include also
           the resolved $\gamma_p$ and the twice resolved
           contributions (a+b+c+d).
           }
\label{Fig2}
\end{figure}

\begin{table}[t]
\caption{Integrated cross section for ZEUS and H1 from
         $p_{\rm t} = 4$ GeV to $p_{\rm t} = 12$ GeV}
\begin{tabular}{|c|c|c|c|c|}
\null\hfil & (a) & (b) & (c) & (d) \\ \hline
ZEUS & 9.57 pb & 0.75 pb & 2.03 pb & 1.38 pb \\ \hline
H1   & 2.92 pb & 0.29 pb & 0.71 pb & 0.59 pb 
\end{tabular}
\end{table}

\section{Photon-Parton Processes in Hadron Colliders}
\begin{figure}[htbp]
\begin{center}
\leavevmode
{
 \epsfxsize=3.00in
 \epsfbox{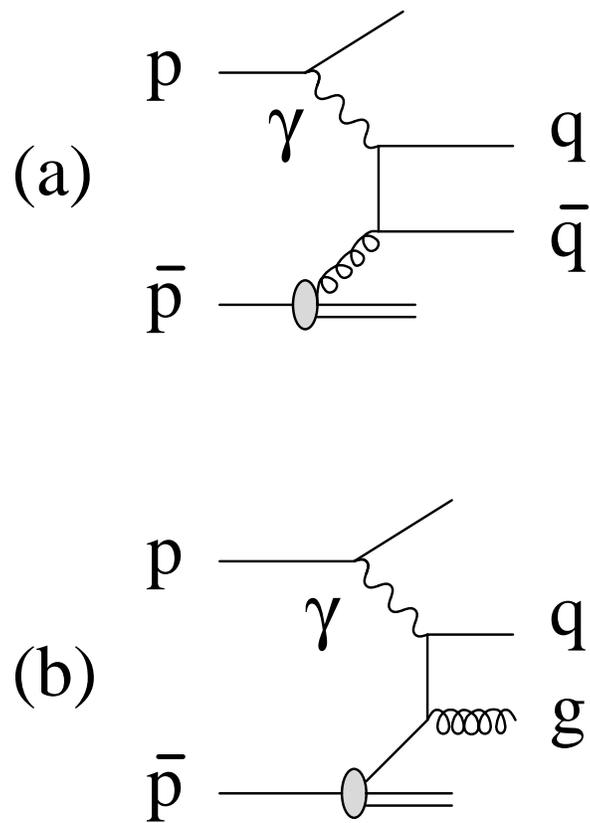}
}
\end{center}
\caption[*]{
           Photon-parton jet processes at $p\bar{p}$ collider.
           (a) photon-gluon fusion, 
           (b) ``Compton" scattering. 
           }
\label{Fig3}
\end{figure}

In this section we analyze the direct photon
contribution to two-jet events at the CERN
$Sp\bar{p}S$ collider.    
The scattering mechanisms are shown in Fig. 3.   
For concreteness, we assume
that the direct photon originates from the proton,
although experimentally the conjugate reaction
with photon coming from the antiproton is also
accounted.  We recall that it was in these events that the superhard 
pomeron was first reported\cite{UA8}.   
Having shown that direct photons are the leading twist contribution 
to the superhard dijet diffractive events at $ep$ colliders,  it is thus 
natural to speculate if the UA8 events were not simply due to direct photons.  
This speculation is further fed by the fact that hadronic mechanisms 
are once again higher--twist.

\subsection{Hadronic Mechanism is Higher Twist}
\begin{figure}[htbp]
\begin{center}
\leavevmode
 {
 \epsfxsize=6.00in
 \epsfbox{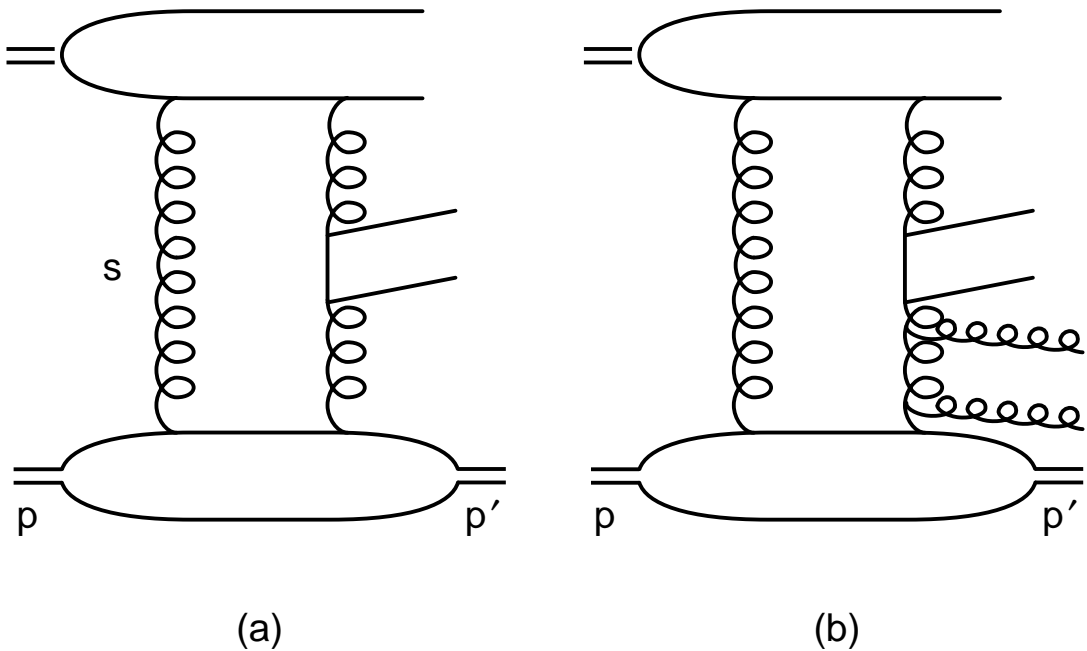}
 }
\end{center}
\caption[*]{
           (a)The simplest mechanism for diffractive jet production in
hadron collisions.   
            The gluon labeled `s' is an arbitrarily soft gluon.  This
mechanism however does 
            not lead to the superhard events seen at UA8 due to gluon
bremsstrahlung, 
           shown in (b).
           }
\label{Fig4}
\end{figure}

The simplest hadronic mechanism for diffractively producing two jets 
at hadronic colliders is shown in Fig. 4a.   The gluon labeled ``s'' is an 
arbitrarily soft gluon.  It's sole purpose is to guarantee that the proton 
is reformed as a color singlet.  As discussed in Ref.
\cite{CollinsFrankfurtStrikman}, 
the  unitarity cancellation over all such soft gluon attachments that occurs 
in the case of inclusive processes cannot be implemented due to the 
constraint that a proton is selected to be found in the final state.    
Nevertheless, the contribution of this hadronic mechanism to the 
superhard events seen by UA8 is asymptotically ignorable due to Sudakov 
suppression.   

Unless constrained otherwise, the hot gluon that 
enters the hard interaction will radiate, as shown in 
Fig. 4b.   Such radiation will appear in the final state as momentum not associated 
with the dijet system.  Hence it will not contribute to the ``superhard'' 
pomeron events seen by UA8 wherein all the energy lost by the diffracted 
proton appears in the dijet system.  

To prevent the radiation depicted in Fig. 4b from occurring, the soft gluon 
(labeled ``s'') must be drawn into the hard interaction.   This introduces a 
Sudakov form factor\cite{Mueller} which amounts to an effective 
fractional power of  suppression of the hadronic mechanism, i.e. 
$(\Lambda_{QCD}/E_\perp)^{\alpha(E_\perp)}$.  
In the case of elastic scattering of protons at high 
momentum transfers, it is known that for uncorrelated 
quark scattering (the so--called Landshoff mechanism\cite{Landshoff}) 
$\alpha(E_\perp) \approx 1$\cite{BottsSter}.   For the present case   
$\alpha(E_\perp)$ has not been calculated\footnote{And indeed represents one 
of the most important open theoretical problems in the field of 
hard diffractive scattering.}.     However one can nevertheless conclude that 
due to these Sudakov effects, the hadronic mechanism is higher--twist and 
hence that asymptotically (i.e. for fixed $t$ and fixed fractional 
energy loss of the proton) direct photons will also dominate 
the superhard pomeron type events as reported by UA8.   
The extension to double--diffractive hard events is obvious.

The fact that in hadronic colliders, the hadronic mechanism is 
not suppressed by clear full power of $1/E_\perp$, in 
contradistinction to the case for HERA, is a 
manifestation that in this class of events the hadronic mechanism does not 
factorize\cite{CollinsFrankfurtStrikman}. From 
the above arguments and those of Section (IIa), 
we see that factorization is in fact broken at the higher--twist level.  
It becomes now a quantitative question whether in any particular process  
the overall numerical factors  of $\alpha_{\rm em}$ suppresses the 
leading twist, direct photon 
mechanism.  We now address this question in the case of the UA8 data.

\subsection{Direct Photons at UA8}

The cross section for the direct photon contribution
to two-jet events, as shown by the mechanisms in Fig. 3,
is given by
\begin{equation}
d\sigma_{p\bar{p}}
=
f_{\gamma|p} (x_1) 
\left[ \sum_{q, \bar{q}} f_{q|\bar{p}} (x_2,\sim 4 p_{\rm t}^2)
        d \sigma_{q\gamma} (x_1 x_2 s)
+
      f_{g|\bar{p}} (x_2,\sim 4 p_{\rm t}^2)
        d \sigma_{g\gamma} (x_1 x_2 s)
\right] dx_1 dx_2,\label{cross1}
\end{equation}
where the cross section $d\sigma_{q\gamma}=d\sigma_{\bar{q}\gamma}$
and $d\sigma_{g \gamma}$ are as given in formulas
(\ref{QuarkPhotonCrossSection}) and 
(\ref{GluonPhotonCrossSection}).
Note that the we've taken the scale of the quark and gluon distribution amplitudes 
at the total invariant mass of the dijet system.   We will return to
the sensitivity 
of our results upon this choice.  

We can re-arrange Eq. (\ref{cross1}) into
\begin{equation}
d\sigma_{p\bar{p}}
=
\frac{2\pi \alpha_{\rm em} \alpha_{\rm s} }
     {x_1 x_2^2 s}
f_{\gamma|p} (x_1)
\left[ \frac{5+2 z + z^2}{3 (1+z)} F_2(x_2,4 p_{\rm t}^2)
+ \frac{1+z^2}{1-z^2}
\left( \sum_q Q_q^2
\right)
x_2 g(x_2,4 p_{\rm t}^2)
\right] dx_1 dx_2 dz.
\label{crossUA8}
\end{equation}
For the distribution function $F_2(x_2,4p_{\rm t}^2) = 
\sum_{q, \bar{q}} Q_q^2 x_2 f_{q|p} (x_2,4p_{\rm t}^2)$ 
we use the recent parametrization\cite{ZEUSMeasurement} given by 
the ZEUS collaboration of their data
\begin{equation}
F_2(x,Q^2) = \left(1 - x^2 \right)^4
\left[ .35 + .017  x^{-(.35+.0695{\rm ln} Q^2)}\right].
\end{equation}
For the gluon momentum density $x_2 g(x_2,4p_{\rm t}^2)$ 
we use the low-$x$ approximation scheme of Prytz\cite{Prytz} to the evolution 
equations whereby one relates, at leading order, 
\footnote{We obtain similar numerical results if we 
use the scheme of EKL\cite{EKL} and take $\omega_0 = .4$.  We also obtain 
similar numerical results for $x g(x,Q^2)$ for  $Q^2 \approx 64 $GeV$^2$ 
if we used the H1 parametrization \cite{H1F2Measurement}  of their data.  
However this latter parametrization does not apply at larger $Q^2$ for the 
low $x_2$ values ($\lesssim .01$) needed here.} 
\begin{equation}
x g(x,Q^2) = \frac{27 \pi}{10 \alpha_s(Q^2)} \frac{d F_2(x/2,Q^2)}{d {\rm ln}Q^2}.
\end{equation}

The proton's momentum transfer squared at the UA8 experiment
is in the range $|t|=0.9 \sim 2.3$ GeV$^2$.    
For direct photon to alone be responsible for the diffraction of the proton, 
these are the limits that must enter the photon's structure function, 
\begin{eqnarray}
f_{\gamma|p}(x_2)
&=&
\frac{\alpha_{\rm em}}
     {2\pi}
\int_{.9 \ {\rm GeV}^2}^{2.3 \ {\rm GeV}^2}
\frac{d Q^2}
     { Q^2 }
\left[ 2 \left( F_1^2(Q^2) + \frac{Q^2}{4 m_p^2} F_2^2(Q^2) \right) \times \right.\\
\nonumber
&&\left. \left(  \frac{1-x_2 }{x_2}-  \frac{m_p^2}{Q^2} x_2 \right) 
+ \left( F_1(Q^2) + F_2(Q^2) \right)^2 x_2  \right].
\label{fphinptrun}
\end{eqnarray}

In Table III we present the integrated cross section with
the kinematic cuts of the UA8 group. That is, we demand
that the hard-jet cone-center rapidities be restricted to the interval
$[-2,2]$, that $0.04 < x_1 < 0.10$, and that $p_{\rm t} > 8$ GeV.
 From the table, we see that the direct--photons  can
account for  $18$\% percent  of the observed 
``super-hard" pomeron events.     The most sensitive  unknown quantity 
in this result is the gluon's structure function.  Had we for example 
chosen the scale in $x g(x, Q^2)$ to be $p_\perp^2$ instead of $4 p_\perp^2$, 
we would have obtained that direct photons contribute $12$\% of the 
observed events.  In either case, the relative fraction is significant.

\begin{table}[t]
\caption{Comparison of the experimental ``Super-hard" pomeron events
         to direct-photon contributions}
\begin{tabular}{|c|c|c||c|c|} 
              & back-to-back & single-side & direct-photon & expected      \\
              & two-jet      & super-hard  & cross section & direct-photon \\ 
$x_p = 1-x_1$ & events       &  events     &  [pb]         & events        \\ 
 & & ($\times 50\% \times 30\%$) & & ($\times 0.423$ [pb$^{-1}$]) \\ \hline
0.90 --- 0.92 & 77  & 11.6 &  5.1 & 2.1  \\ 
0.92 --- 0.94 & 86  & 12.9 &  5.2 & 2.2  \\ 
0.94 --- 0.96 & 86  & 12.9 &  5.4& 2.3  \\ \hline
    total     & 249 & 37.4 & 15.7 & 6.6 \\ 
\end{tabular}
\end{table}

\section{Conclusion}

Direct photons have been shown to be the leading twist mechanism for diffractive 
dijet production  in which the jets carry away all of the momentum lost by 
the diffracted proton for either electron initiated or hadron initiated events.   
Asymptotically, they dominate all hadronic mechanisms.  

In $ep$ collisions, direct photons are predicted to asymptotically dominate 
all such ``superhard'' pomeron, dijet events.    At the present time the inability 
at HERA\cite{ZeusPhotoproduction,H1Photoproduction} to determine the final state 
proton's momentum  
means no data on this particular  class of hard diffraction events yet exists. 
While we wish to strongly stress the importance that such experiments
be undertaken, 
we also want to suggest an alternative experiment that should indirectly 
indicate the relative importance of direct photons, namely, large rapidity gap, 
dilepton production at large transverse momentum.\footnote{We thank 
Shmuel Nussinov for this suggestion. See also \cite{Levman},
where muon-pair production at HERA by the two-photon mechanism
has been studied.} 
 From Eq.~(\ref{direct}) the ratio of dijets to dileptons at the same 
large transverse momentum is simply
\begin{equation}
R_{\gamma \gamma} 
= \frac{N_c \sum_q Q_q^4 }{\sum_l Q_l^4}
= \frac{34}{81}
\end{equation}
where we have summed over four light-quark flavors ($u,d,s,c$) 
and included the $\tau$-lepton in the sum over leptons. $N_c = 3$ is
the number of colors.
We note that to the extent that resolved photon contributions (i.e. as
in Fig. (1b)) 
can be excluded, such dilepton events are absent in a convolution  picture  
of the pomeron\cite{IngelmanSchlein}. 

In $p\bar{p}$ collisions, we have seen  that direct-photons 
 account for an appreciable fraction of the ``super-hard" pomeron events observed 
by UA8\cite{UA8}  at the CERN $Sp\bar{p}S$ collider.    The moderately 
large values of $t$ in these experiments is both playing a significant
role in suppressing the 
direct-photon contribution (via the proton's Dirac form factor) as well as 
what must be inferred as providing a significant scale to partially
limit the effects 
of Sudakov radiation on the competing hadronic mechanism.   It should be 
emphasized that in a complete calculation the hadronic mechanism must be added 
to the direct-photon's contribution to obtain the $S$--matrix amplitude.  The 
relative size of the two contributions could be much closer in magnitude than 
even the above result indicates, depending upon their relative phase.  Interesting 
interference effects are thus likely if either $t$ or the dijet transverse momentum 
could be systematically varied.  Such capabilities will hopefully be realized 
at the set of experiments presently being conducted at FNAL.

\acknowledgments
We also thank J. C. Collins, N. C. Mukhopadhyay, N. I. Sarcevic for discussions, 
H. Abramowicz for useful correspondence and J. Zsembery for providing
us with the UA8 luminosity information. 
This work was supported in part by DOE Grant 
DOE--FG02--93ER--40762 and DOE--FG03--93ER--40792.


\begin{references}

\bibitem{ZeusDIS}
ZEUS Collaboration (M. Derrick et al.),
Phys. Lett. {\bf B315}, 481 (1993),
{\it ibid} {\bf 332}, 228 (1994).

\bibitem{H1DIS}
H1 Collaboration (T. Ahmed et al.),
Nucl. Phys. {\bf B429}, 477 (1994).

\bibitem{ZeusPhotoproduction}
ZEUS Collaboration (M. Derrick et al.),
Phys. Lett. {\bf B346}, 399 (1995),
preprint DESY-95-115.

\bibitem{H1Photoproduction}
H1 Collaboration (T. Ahmed et al.),
Nucl. Phys. {\bf B435}, 3 (1995).

\bibitem{IngelmanSchlein} 
G. Ingelman and P. Schlein,  
Phys. Lett. {\bf B152}, 256 (1985).

\bibitem{CollinsFrankfurtStrikman}
J.C. Collins, L. Frankfurt and M. Strikman, 
Phys. Lett. {\bf B307}, 161 (1993).

\bibitem{BerSop}
A. Berera and D. E. Soper, 
Phys. Rev. {\bf D50}, 4328 (1994).

\bibitem{LuMilana}
H.J. Lu and J. Milana, 
Phys. Lett. {\bf B313}, 234 (1993).

\bibitem{DL2} 
A. Donnachie and P. V. Landshoff,  
Phys. Lett. {\bf B285}, 172 (1992).

\bibitem{UA8} 
UA8 Collaboration, 
Phys. Lett. {\bf B297}, 417 (1992).

\bibitem{Mueller}
A. H. Mueller, 
Phys. Lett. {\bf B108}, 355 (1982).

\bibitem{BottsSter}
J. Botts and G. Sterman,
Nucl. Phys. {\bf B325}, 62 (1989).

\bibitem{LuM}
H.J. Lu and J. Milana, 
Phys. Rev. {\bf D51}, 6107 (1995).  
Only the manifestly short--distance graphs were however included 
in the estimates reported here.  The necessity of including 
additional ``long--distance'' processes with Sudakov suppression 
factors complicates some of the claims made concerning factorization.

\bibitem{BialasLandshoff}
A. Bia\l as and P.V. Landshoff,
Phys. Lett. {\bf B256}, 540 (1991).

\bibitem{Pumplin}
J. Pumplin, 
``Two Gluon Exchange Model Predictions for Double Pomeron Jet Production,''
 hep--ph/9412381,  Michigan State University preprint MSUHEP--41222 (1994).

\bibitem{BerCollins}
A. Berera and J. C. Collins, 
``Double Pomeron Jet Cross Sections,'' 
 hep-ph/9509258, Pennsylvania State University preprint PSU/TH/162 (1995).
These authors state however that they expect large Sudakov 
corrections to their estimates.

\bibitem{DreesGodbolePramana}
M. Drees and R. M. Godbole,
Pramana J. Phys. {\bf 41}, 83 (1993).

\bibitem{WeizsackerWilliams} 
C. F. Weizs\"acker, Z. Phys. {\bf 88}, 612 (1934), 
E. J. Williams Phys. Rev. {\bf 45}, 729 (1934).

\bibitem{EETwoPhoton}
S. J. Brodsky, T. Kinoshita and H. Terazawa,
Phys. Rev. {\bf D4}, 1532, (1971).
R. Bhattacharya, J. Smith and G. Grammer, Jr.
Phys. Rev. {\bf D15}, 3267 (1977).

\bibitem{Bussey}
P.J. Bussey,
in Proceedings, Physics at HERA, vol. {\bf 1} (1991).

\bibitem{GamGamHiggs}
E. Papageorgiu, Phys. Rev. D{\bf 40}, 92 (1989); 
M. Drees, J. Ellis, and D. Zeppenfeld, 
Phys. Lett. {\bf B223}, 454 (1989); 
M. Grabiak, B. M\"uller, W. Greiner, B. Soff, and P. Kock, 
J. Phys. {\bf G15}, L25 (1989);  
R. N. Cahn and J. D. Jackson,
Phys. Rev. {\bf D42}, 3691 (1990); 
J. Norbury, 
Phys. Rev. {\bf D42}, 3696 (1990); 
B. M\"uller and A. J. Schramm 
Phys. Rev. {\bf D42}, 3699 (1990).

\bibitem{ffdata}
L. Andivahis {\it et al.},
Phys. Rev. {\bf D50}, 5491 (1994).

\bibitem{Nimai}
R. M. Davidson,  N. C. Mukhopadhyay, and R. S. Wittman, 
Phys. Rev. {\bf D43}, 71 (1991).    

\bibitem{BargerPhillips}
V. D. Barger and R. J. N. Phillips,
Collider Physics, Addison-Wesley Publishing Co. (1987).

\bibitem{Landshoff}
P.V. Landshoff, Phys. Rev. D{\bf 10}, 1024 (1974).

\bibitem{ZEUSMeasurement} 
ZEUS Collaboration (M. Derrick et al.), 
Zeitschrift fur Physik {\bf C65}, 379 (1995).

\bibitem{Prytz}
K. Prytz, Phys. Lett. {\bf B311}, 286 (1993).

\bibitem{EKL}
R. K. Ellis, Z. Kunszt and E. M. Levin, 
Nucl. Phys. {\bf 420}, 517 (1994).

\bibitem{H1F2Measurement}
H1 Collaboration (T. Ahmed et al.),
Nucl. Phys. {\bf B 439}, 471 (1995).

\bibitem{Levman}
G. Levman,
in Proceedings, Physics at HERA, vol. {\bf 1} (1991).

\end{references}
\end{document}